%% file: ms_pp.tex
\documentclass{article}
\usepackage{times}
\usepackage[onecolumn]{emulateapj}
\usepackage{natbib}

\usepackage{flushrt}
\input aas_journals.tex

\newcommand{\ltsima}{$\; \buildrel < \over \sim \;$}
\newcommand{\lsim}{\lower.5ex\hbox{\ltsima}}
\newcommand{\gtsima}{$\; \buildrel > \over \sim \;$}
\newcommand{\gsim}{\lower.5ex\hbox{\gtsima}}

\def\gtrsim{\mathrel{\hbox{\rlap{\hbox{\lower4pt\hbox{$\sim$}}}\hbox{$>$}}}}

\def\lesssim{\mathrel{\hbox{\rlap{\hbox{\lower4pt\hbox{$\sim$}}}\hbox{$<$}}}}

\begin{document}
\title{Wide Angle Redshift Distortions Revisited}

\author{Istv\'an Szapudi \footnote{ Institute for Astronomy, 
University of Hawaii, 2680 
Woodlawn Dr, Honolulu, HI 96822} 
}

\begin{abstract}

We explore linear redshift distortions in wide angle
surveys from the point of view of symmetries.
We show that the redshift space two-point correlation function can
be expanded into tripolar spherical harmonics of zero total angular momentum 
$S_{l_1 l_2 l_3}(\hat x_1, \hat x_2, \hat x)$. 
The coefficients of the expansion $B_{l_1 l_2 l_3}$
are analogous to the $C_l$'s of
the angular power spectrum, and express the anisotropy of the
redshift space correlation function. Moreover, 
only a handful of $B_{l_1 l_2 l_3}$ are non-zero:
the resulting formulae reveal a hidden simplicity comparable to
distant observer limit. The $B_{l_1 l_2 l_3}$   
depend on spherical Bessel moments of the power spectrum 
and $f = \Omega^{0.6}/b$. 
In the plane parallel limit, the results of \cite{Kaiser1987} and
\cite{Hamilton1993} are
recovered. The general formalism is used to derive 
useful new expressions. We present a particularly simple
trigonometric polynomial expansion, which 
is arguably the most compact expression of wide angle redshift distortions.
These formulae are suitable to inversion due to the
orthogonality of the basis functions. An alternative
Legendre polynomial expansion was obtained as well. This can be
shown to be equivalent to the results of \cite{SzalayEtal1998}.
The simplicity of the underlying theory will admit similar
calculations for higher order statistics as well.

\end{abstract}

\section{Introduction}

It has been known for decades that the two-point correlation
function, or power spectrum, measured in redshift surveys
is distorted by the peculiar velocities of galaxies. The anisotropy
of the correlation function was demonstrated by 
\cite{DavisPeebles1983,Peebles1980}. In a seminal work,
\cite{Kaiser1987} demonstrated that in the plane parallel limit
there is a simple transformation between the redshift space and
real space density contrast. This  results in an anisotropic enhancement
of the power spectrum by $(1+f\mu^2)^2$, where $\mu$ is the
cosine of the angle between the line of sight and the wave-vector.
This simple formula has become the starting point of many extensions,
which have used expansion into Legendre polynomials 
\citep[e.g.,][]{Hamilton1993,HamiltonCulhane1996},
or numerical methods \citep{ZaroubiHoffman1996}.
Most analyses assume a small opening angle \citep{ColeEtal1995}, 
i.e. they stay essentially in the distant observer limit.
Others works used 
a expansion with formally infinite number of coefficients 
\citep{HeavensTaylor1995}. Numerous galaxy redshift surveys
have been successfully analyzed with such methods, most notably
the PSCz \cite{TadrosEtal1999}, 
2dF \citep{PeacockEtal2001,HawkinsEtal2003, TegmarkEtal2002}, 
and SDSS \citep{ZehaviEtal2002}. For a review of methods in the above spirit
and the corresponding applications, see \cite{Hamilton1998}.

To address the needs of wide angle redshift surveys, 
full treatment of wide angle distortions have been given by
\cite{SzalayEtal1998}, where they identify the coordinates
in which the expression of the redshift space two-point
correlation function is compact, and most importantly finite.
They have also argued,
that the power spectrum will necessarily have an infinite expansion,
as it arises from the convolution of the density field 
with a non-compact kernel.
They concluded that correlation functions are more convenient 
for redshift space analyses then power spectra.
Their results is suitable and has been used for ``forward'' analyses,
such as the Karhunen-Loeve method, in which the correlation function
is predicted and contrasted with data. Applications
to the SDSS are presented most recently by \cite{PopeEtal2004}, 
\citep[see also][]{TegmarEtal2003a,TegmarEtal2003b}.

In this paper we analyze the symmetries of redshift space
distortions. The next section shows that zero angular momentum
tripolar functions form a natural basis to expand the redshift
space correlation function, and that only a surprisingly 
small set of expansion coefficients
are will be non-zero. In section 3 we present the most important properties
of the basis functions, and the connection with the Kaiser-Hamilton limit.
Section 4 employs the general theory to obtain compact expressions
for the redshift distortions using conveniently chosen variables.
In the final section we present discussions, and conclusions.

\section{Redshift Distortions in Linear Theory}

The theory of redshift distortions is based on the 
redshift to real space transformation, $s_i = x_i - f v_j \hat x_i \hat x_j$,
where $\hat x_i$ is a unit vector pointing to the galaxy from the origin,
$f = \Omega^{0.6}/b$,
and $v_j$ is the peculiar velocity in units that its divergence equal
to the density to linear order \citep{ScoccimarroEtal1999}.
The transformation of the density $\delta$ then can be estimated via
the linear theory Jacobian
$J = 1-f \hat x_i \hat x_j \partial_j v_i$, as
\begin{equation}
  \delta_s(x) = \int \frac{d^3 k}{(2\pi)^3}e^{ik_sx_s}
   (1+f(\hat x_s \hat k_s)^2)
  \delta(k)
\end{equation}
where the Einstein convention was followed for summation of multiple
indices. The effects due to the gradient of the selection
function and the ``rocket effect'' \citep{Kaiser1987} were neglected.
As a consequence, the redshift distorted 2-point correlation
function $\xi^s(x_1,x_2)$ reads
\begin{eqnarray}
  \xi^s(\hat x_1,\hat x_2, \hat x, x) & = (1+\frac{1}{3}f)^2 \xi(x)\cr
   &+ (\frac{2}{3}f+\frac{2}{9}f^2) \int \frac{d^3 k}{(2\pi)^3}e^{ik \hat x x}
   P_2(\hat k \hat x_1) P(k) + (1 \leftrightarrow 2)^* \cr
   &+ \frac{4}{9}f^2\int \frac{d^3 k}{(2\pi)^3}e^{ik \hat x x}  
   P_2(\hat k \hat x_1)P_2(\hat k \hat x_2)P(k),
\label{eq:xis}
\end{eqnarray}
where $P_2$ is the second order Legendre polynomial, $P(k)$ is the
linear power spectrum. We have chosen to characterize the correlation
with the two directions from the line of
sight, and introduced $x = x_1 - x_2$, and the corresponding unit
vector. By construction, these three unit vectors are in the same plane.
The above formula is in full agreement with \cite{SzalayEtal1998}.

A few simple observations are in order with respect to the above
function i) it is rotationally invariant ii)
$\xi^s(\hat x_1,\hat x_2, -\hat x, x)=\xi^{s*}(\hat x_1,\hat x_2, \hat x, x)=
\xi^s(\hat x_1,\hat x_2, \hat x, x)$, since the correlation function is real, 
iii)  $\xi^s(\hat x_1,\hat x_2, \hat x, x)=\xi^s(\hat x_2,\hat x_1, \hat x, x)$
the invariance of the correlation function under permutation,
iv) the three
vectors are constrained to be in the same plane, v) the unit vector
$\hat x$ is constrained to be between $\hat x_1$ and $-\hat x_2$.
vi) We can extend the function for $\hat x$ vectors outside
this range, with the definition 
$\xi^s(\hat x_1,\hat x_2, \hat x, x)\equiv 
\xi^s(\hat x_1,-\hat x_2, \hat x, x)$.
This leaves Equation~\ref{eq:xis} formally valid, 
since $P_l(-\mu) = (-1)^lP_l(\mu)$,
and the equation contains only even Legendre polynomials.

For a function depending on three directions, the tripolar
spherical harmonics expansion \citep[e.g., ][]{VarshalovichEtal1988}
is the most natural
\begin{eqnarray}
  \xi^s(\hat x_1,\hat x_2, \hat x, x) & = \sum B^{l_1 l_2 l_3 l_{23}}_{LM}(x)
(Y_{l_1} \otimes (Y_{l_2}\otimes Y_{l_3})_{l_{23}})_{LM}\cr
 &=\sum B^{l_1 l_2 l_3 l_{23}}_{LM}(x)D^L_{M M^\prime}
(Y_{l_1} \otimes  (Y_{l_2}\otimes Y_{l_3})_{l_{23}})_{LM^\prime},
\end{eqnarray}
where $D^L_{M M^\prime}$ is the Wigner rotation matrix, and
the second line follows from rotational invariance. As a consequence 
{\em only} $L=0$, i.e. zero total 
angular momentum, can appear in the
above expansion. This means that quite generally, the expansion
will only contain the functions  
$(Y_{l_1} \otimes (Y_{l_2}\otimes Y_{l_3})_{l_{23}})_{00}$ with
$l_{23}=l_1$. In order
to simplify the final expressions, we will use the following
functions, which are proportional to the zero angular momentum
tripolar functions
\begin{equation}
S_{l_1 l_2 l}(\hat x_1,\hat x_2, \hat x) \equiv
\sum_{m_1,m_2,m}
\left( \begin{array} {ccc} l_1 & l_2 & l \\ m_1 & m_2 & m
             \end{array} \right) 
      C_{l_1 m_1}(\hat x_1) C_{l_2 m_2}(\hat x_2) C_{l m}(\hat x)
\end{equation}
where $\left( \begin{array} {ccc} l_1 & l_2 & l \\ m_1 & m_2 & m
             \end{array} \right)$ is the Wigner $3j$ symbol, and
$C_{lm} = \sqrt{4\pi/2l+1}Y_{lm}$ are proportional to the spherical
functions. 

Expanding Equation~\ref{eq:xis}, and using 
$P_l(\hat x_1\hat x_2)=\sum_m C_{lm}(\hat x_1)C^*_{lm}(\hat x_2)$,
$e^{ik x}=\sum_{lm}(2l+1)i^lj_l(kx) C_{lm}(\hat k)C^*_{lm}(\hat x)$,
and the Gaunt integral, one find finds that
\begin{equation}
  B^{l_1 l_2 l l_1}_{00} \equiv B^{l_1 l_2 l} = 
   (2l+1)\xi_l(x) i^l 
   \left( \begin{array} {ccc} l_1 & l_2 & l \\ 0 & 0 & 0
             \end{array} \right) F_{l_1}F_{l_2},
\end{equation}
where we have introduced 
$\xi_l(x) = \int \frac{dk}{2\pi^2} k^2 j_l(x k) P(k)$, the moments of the
power spectrum with spherical Bessel functions, and
$F_0 = 1+1/3f,\,F_2=2/3f$, otherwise 0. 
From property (iii) it follows
that $B^{l_1 l_2 l_3}=B^{l_2 l_1 l_3}$

It is worth to write
this result explicitly, since only a few
terms are non-zero due to the initial expression and constraints
from group theory: 
\begin{eqnarray}
  B^{000}(x) &= (1+\frac{1}{3}f)^2 \xi_0(x)\cr
  B^{220}(x) &=\frac{4}{9\sqrt{5}}f^2 \xi_0(x)\cr
  B^{022}(x)= B^{202}(x) &= 
  -(\frac{2}{3}f+\frac{2}{9}f^2)\sqrt{5} \xi_2(x)\cr 
  B^{222}(x) &= \frac{4\sqrt{10}}{9\sqrt{7}}f^2\xi_2(x)\cr
  B^{224}(x) &= \frac{4\sqrt{2}}{\sqrt{35}}f^2\xi_4(x).
  \label{eq:main}
\end{eqnarray}
These functions, not unlike the $C_l$ for the angular power spectrum,
form a natural basis for maximum likelihood estimation. They
can be used as an intermediate step for  estimation of 
cosmological parameters in the linear regime.

\section{Properties of the $S_{l_1 l_2 l_3}$ functions}

The  $S_{l_1 l_2 l_3}$ functions are a subset of tripolar spherical
harmonics. They form an orthogonal complete basis for expanding
spherically symmetric functions depending on three unit vectors.
It should be emphasized, that  orthogonality is true only
 when  the unit vectors are  integrated over the full sphere
$d\Omega_1 d\Omega_2 d\Omega_3$ unrestricted. From the definition
it follows that 
\begin{equation}
  \int d\Omega_1 d\Omega_2 d\Omega_3
 S_{l_1 l_2 l_3}(\hat x_1,\hat x_2, \hat x)
 S_{j_1 j_2 j_3}(\hat x_1,\hat x_2, \hat x) = \delta_{l_1 j_1}
\delta_{l_2 j_2} \delta_{l_3 j_3} 
 \frac{(4\pi)^{3/2}}{\sqrt{(2l_1+1)(2l_2+1)(2l_3+1)}}.
\label{eq:ortho}
\end{equation}

The plane parallel limit \citep{Kaiser1987} can be obtained by
assuming that the first two unit vectors are parallel. Using the
properties of the Wigner coefficients and spherical functions, 
it is easy to show that
\begin{equation}
  S_{l_1 l_2 l_3}(\hat x_1,\hat x_1, \hat x)
   =  \left( \begin{array} {ccc} l_1 & l_2 & l_3 \\ 0 & 0 & 0
             \end{array} \right) P_{l_3}(\hat x_1 \cdot \hat x).
\end{equation}
As a special case one can see that independently of the
unit vector  $\hat x_1$
\begin{eqnarray}
  S_{0 0 0} & = 1\cr
  S_{0 l l}(\hat x_1,\hat x_2) = S_{l 0 l}(\hat x_1,\hat x_2) =
  S_{l l 0}(\hat x_1,\hat x_2)
   &= \frac{(-1)^l}{\sqrt{2l+1}}
  P_l(\hat x_1 \cdot \hat x_2) \cr
\end{eqnarray}
and all other functions with any zero index are zero.

It follows from the above properties by simple algebra that
in the plane parallel limit $\hat x_1 \simeq \hat x_2$
\begin{eqnarray}
  \xi^s(\hat x_1,\hat x_1, \hat x, x) 
   &= (1+\frac{2}{3}f+\frac{1}{5}f^2)\xi_0 +\cr
                 & -(\frac{4}{3}f+\frac{4}{7}f^2)\xi_2 P_2(1,3)\cr
                 &+\frac{8}{35}f^2\xi_4 P_4(1,3),
\end{eqnarray}
in agreement with \cite{Kaiser1987,Hamilton1993,HamiltonCulhane1996}.

\section{Coordinate Systems}

Since the three unit vectors are constrained in the same
plane (property (iv), the redshift space correlation function
depends only on the shape and size of a triangle. Consequently,
the angular dependence can be parametrized by two angles.
Since the  $S_{l_1 l_2 l}$ functions are rotationally
invariant, we can fix the plane of the vectors, and even rotate
one of them to a fixed position in order
to obtain useful expressions. We explored the following choices 
A) the z axis is perpendicular
to the plane of the vectors, and we fix $\phi=0$ for $\hat x$, B) 
z axis coincides with the third unit vector, $\hat z = \hat x$,
C) it points in between the first two unit vectors 
$\hat z \propto \hat x_1 + \hat x_2$. For the latter choices,
we can assume that all the vectors are in the $\phi=0$ plane.

For choice A), $S_{l_1 l_2 l}(\hat x_1,\hat x_2, \hat x) = 
S_{l_1 l_2 l}(\pi/2,\phi_1,\pi/2,\phi_2,\pi/2,0)$, and one obtains
an expansion of the form 
$\xi_s(\phi_1,\phi_2,x) = \sum_{n_1,n_2 = 0,2} a_{n_1 n_2} 
\cos(n_1 \phi_1)\cos(n_2 \phi_2)+ b_{n_1 n_2} 
\sin(n_1 \phi_1)\sin(n_2 \phi_2)$, where the only non-zero
coefficients are:
\begin{eqnarray}
a_{00 }&=
\left( 1 + \frac{2\,f}{3} + \frac{2\,f^2}{15} \right)
     \,\xi_0(x) - \left( \frac{f}{3} +
     \frac{2\,f^2}{21} \right) \,\xi_2(x) + 
  \frac{3\,f^2}{140}\,\xi_4(x)\cr
a_{02}=a_{20}&= \left( \frac{-f}{2} - 
        \frac{3\,f^2}{14} \right) \,\xi_2(x) + 
     \frac{{f}^2}{28}\,\xi_4(x)  \cr
a_{22}& =  \frac{{f}^2}{15}\,\xi_0(x) - 
     \frac{f^2}{21}\,\xi_2(x) + 
     \frac{19\,f^2}{140}\,\xi_4(x) \cr
b_{22}& = \frac{{f}^2}{15}\,\xi_0(x) - 
     \frac{f^2}{21}\,\xi_2(x) - 
     \frac{4\,f^2}{35}\,\xi_4(x)\cr
   \label{eq:trig}
\end{eqnarray}
According to property (v) there is a restriction that $\phi_1 \le \phi_2$.
which, however, can be lifted by symmetry property (iii).
For a fixed $x$, 
the two angles can span the full range of the integration, if there 
are no restrictions from incomplete sky coverage.
Then, the above becomes  (double) orthogonal expansion, where the
coefficients can be obtained simply by integration, e.g.
$a_{n_1 n_2}\propto \int_0^{\pi/2} d\phi_1d\phi_2 \xi_s(\phi_1,\phi_2,x)
\cos(2 n_1 \phi_1)\cos(2 n_2 \phi_2)$. If the
correlation function is measured and binned according to the above
expression, the four independent coefficients can be obtained
either by numerical integration, or by fit; the latter would
be probably preferable for realistic surveys with incomplete
sky coverage.

The above expression can be inverted easily with a computer algebra
package: the variables
$f, \xi_0,\xi_2,\xi_4$ can be expressed analytically as a function
of the four coefficients. Unfortunately, the analytical expression
is too complicated to list here (it is a solution of a fourth order
polynomial in $f$), but if needed it can be easily obtained with
any computer algebra package, such as {\tt Mathematica}. In practical
applications, however, numerical inversion is expected to be
more robust.

The plane parallel limit is $\phi_1=\phi_2$: reexpressing
the trigonometric functions in Legendre polynomials indeed yields,
after somewhat tedious calculation,
the familiar expression of \cite{Hamilton1993}.

In coordinate system B) 
$S_{l_1 l_2 l}(\hat x_1,\hat x_2, \hat x) = 
  S_{l_1 l_2 l}(\theta_1,0,\theta_2,0,0,0)$, but the meaning
of the two angles is the same as for coordinate system A). It
can be shown with simple but tedious calculation that it reduces to the
same expression as above. 

Finally, coordinate system C) is identical to that of \cite{SzalayEtal1998};
As we show next, it produces a double Legendre expansion. For this choice
we have $S_{l_1 l_2 l}(\hat x_1,\hat x_2, \hat x) = 
S_{l_1 l_2 l}(\theta,0,\theta,\pi,\gamma,0)$
($\phi=\pi$ ensures that the $\hat z$ axis is between the first two
unit vectors). The notation of the variables emphasizes the connection
with \cite{SzalayEtal1998}. 

Using the fact that 
$C_{lm}(\theta, \pi) = (-1)^m C_{lm}(\theta,0)$, inserting the Clebsch-Gordan
expansion 
\begin{equation}
C_{l_1m_1}(\hat x)C_{l_2m_2}(\hat x) = \sum_{l_3 m_3} 
\left( \begin{array} {ccc} l_1 & l_2 & l_3 \\ 0 & 0 & 0
             \end{array} \right)
\left( \begin{array} {ccc} l_1 & l_2 & l_3 \\ m_1 & m_2 & m_3
             \end{array} \right)C_{l_3 m_3}(\hat x),
\end{equation}
and using property (iii), one can show that in $m_3$ and $m$ are both
even. This means that in that case the $S_{l_1 l_2 l}$ functions
have only even associated Legendre functions, which can be reexpressed
into a finite set of Legendre polynomials. Since there is only a few 
non-zero coefficients, we show the explicit result instead
of the general derivation, which is not very illuminating. Finally,
the redshift space two-point correlation function
can be expanded into multiples of two Legendre polynomials,
$\xi_s(\cos\theta\equiv \mu_1,\cos\gamma\equiv \mu_2,x) 
= \sum_{l_1,l_2 = 0,2} c_{l_1 l_2} P_{l_1}(\mu_1)P_{l_2}(\mu_2)$,
where the only non-zero coefficients are
\begin{eqnarray}
  c_{00}= & \left ( 1 + \frac{2\,f}{3} + 
  \frac{29\,f^2}{225}\right )\xi_0(x)  - 
    \left( \frac{4\,f}{9} + 
  \frac{44\,f^2\,}{315} \right)\xi_2(x) + 
  \frac{32\,f^2\,}{1575} \xi_4(x) \cr
c_{02}= &  \left( \frac{4\,f}{9} + 
     \frac{4\,f^2}{21}\right)\xi_2(x) - 
     \frac{8\,f^2}{63}\xi_4(x) \cr
c_{20}= &  -\frac{16\,f^2}{315}\xi_0(x) + 
     \left(\frac{4\,f}{9} + 
     \frac{100\,f^2}{441}\right)\xi_2(x) - 
     \frac{88\,f^2}{2205}\xi_4(x) \cr
c_{22}= & -\left( \frac{16\,f}{9} + 
        \frac{16\,f^2}{21}\right)\xi_2(x) + 
        \frac{8\,f^2 }{63}\xi_4(x)\cr
c_{04}= &  \frac{8\,f^2}{35}\xi_4(x) \cr
c_{40}= & \frac{64\,f^2}{525}\xi_0(x) - 
     \frac{64\,f^2}{735}\xi_2(x) + 
     \frac{24\,f^2}{1225}\xi_4(x)
\label{eq:legendre}
\end{eqnarray}
In this form the plane parallel limit is $\mu_1=1$, i.e.
$P_l(1) = 1$. It is simple (although a bit tedious) 
matter to show that this again returns the right answer. 
Again, simple, tedious calculation shows that our 
expression reproduces the results of \cite{SzalayEtal1998}, 
if in their Eq.~15 the typographical error $4/15 \rightarrow 8/15$ 
is corrected.

The above is formally an orthogonal expansion. 
Due to property (v), 
$\theta \le \gamma \le \pi - \theta$ must be satisfied for
for any given $x$. 
According to property (v), however, the range can be extended,
and the orthogonality of the Legendre polynomials $P_l(\mu_1)$ is ensured.
The coefficients
in the above expansion can be obtained by Gauss-Legendre integration
of the correlation function. Measuring the correlation function
in these coordinates would produce a method which would be the
closest generalization of the original Kaiser-Hamilton method.

The above form perhaps provides the most natural connection
to the plane parallel limit, therefore it can be used to
quantify deviations from it. Figure 3. plots wide angle to
plane parallel ratio of the two most
common estimators for redshift distortions: the ratio
of the redshift space and real space correlation functions,
and Hamilton's $Q(s) = \xi_2(s)/(3/s^3\int\xi_0(y)y^2dy - \xi_0(s))$,
the modified quadrupole to monopole ratio. 
According the Figure,
a simple restriction of the
opening angles at $\theta \lsim 15-20$ degrees would ensure
the accuracy of traditional measurements assuming the
plan-parallel approximation.

\begin{figure}[htb]
\plotone{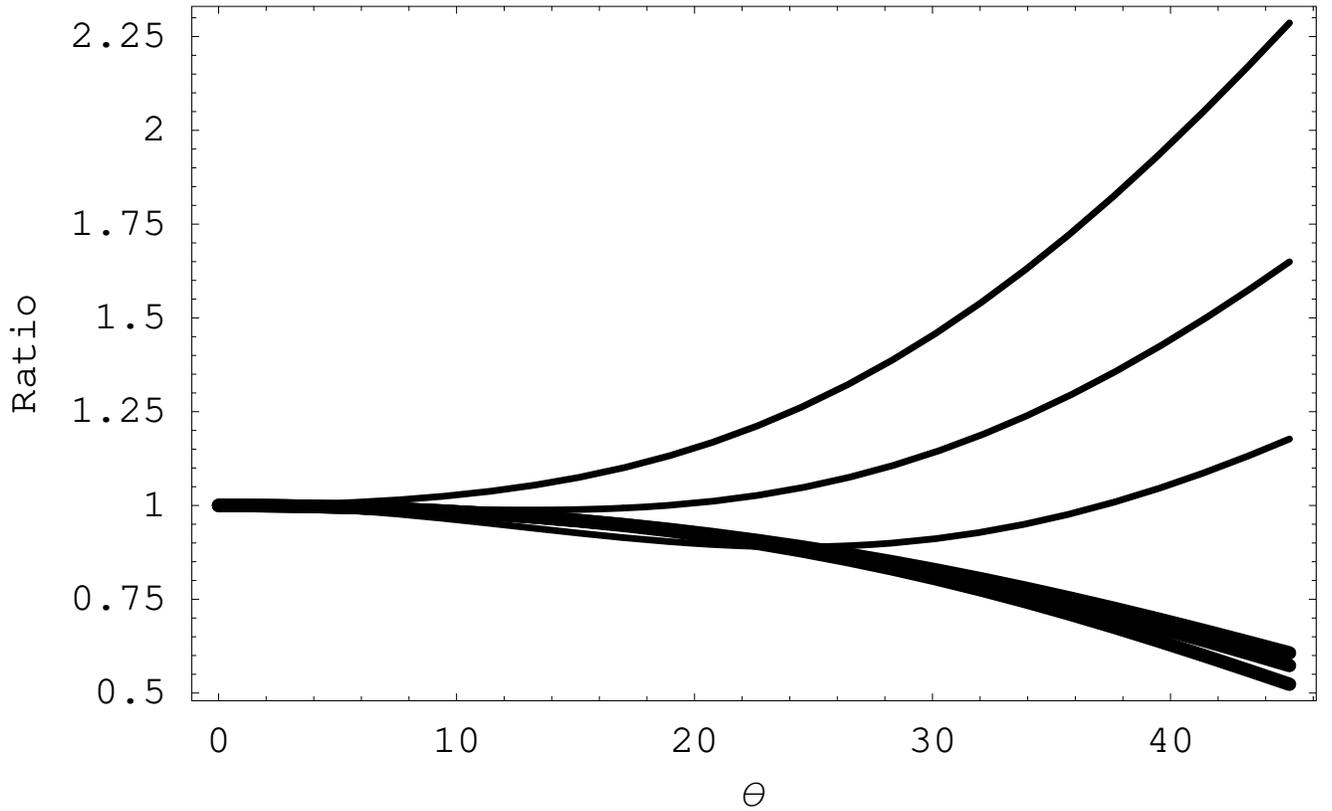}
\caption{The two most common estimators for redshift distortions
are compared to their wide angle analogue: $\xi_0(s)/\xi(r)$ (thick line),
and $Q(s) = \xi_2(s)/(3/s^3\int\xi_0(y)y^2dy - \xi_0(s))$ (thin line). 
The ratio of wide angle to plane parallel prediction is plotted
as a function of the half opening angle $\theta$. The three sets of
lines correspond to three slopes of the correlation
function $1.5,1.75,2.0$, increasing and decreasing for the three
thick and thin lines, respectively.
In a realistic measurement, 
the final result would be a weighted average over a set of
opening angles, represented by the above curves.
}
\end{figure}


\section{Conclusions and Discussion}

We have analyzed the underlying symmetries of redshift space distortions.
As a result, we presented three novel expressions for 
the redshift space two-point correlation functions, 
Eqs.~\ref{eq:main},\ref{eq:trig},and \ref{eq:legendre}. The last of these
equations turns out to be identical to \cite{SzalayEtal1998}.

First, we have shown that the two-point correlation function 
can be expanded into tripolar harmonics and that
rotational symmetry restricts the non-zero components
to those with $L=0$. We have found by direct calculation that
the quadratic nature of linear redshift distortions restricts
the expansion to five unique functions 
$B_{l_1 l_2 l}$ of Equation~\ref{eq:main},
each depending on the distance between the two points only.
These  are analogous to the $C_l$'s of  the angular power spectrum,
constitute a natural basis  for maximum likelihood analysis of redshift space
data. The full machinery of maximum likelihood can be adapted naturally
\cite[e.g.][]{VogelleySzalay1996,TegmarkEtal1998,BondEtal2000}. Details are
left for subsequent research.

The rotational invariance of the basis functions, and the fact
that the three unit vectors are constrained into a plane, 
allows us to fix a plane and an orientation within that
plane. Using this freedom, we identified two convenient
coordinate systems, which correspond to particular choices
of variables. The resulting expressions, 
Eqs.~\ref{eq:trig},and\ref{eq:legendre}
are especially convenient, since they respectively correspond
trigonometric and Legendre polynomial expansions in the two angular variables.
The first expression is possibly most compact formula
for linear redshift space distortions of the two-point correlation function,
the second can be shown to be identical to \cite{SzalayEtal1998}.
The orthogonality in these expansions presents an opportunity
for applications similar in spirit to the original Kaiser-Hamilton method,
but fully correct for wide angles.

This paper deals with the theory of redshift distortions, and thus
lays the groundwork for possible future applications.
We have not discussed practical issues, such as incomplete
sky coverage, noise.  Clearly, even if these issues are important,
a maximum likelihood technique to find the parameters of the
expansions would be still optimal. On the other hand, 
incomplete sky coverage and noise will cause leakage,
possible emergence of higher $l$ anisotropies. We conjecture
that such difficulties for direct methods could be solved similarly to 
\cite{SzapudiEtal2001}, where the analogous problem for the angular
power spectrum was tackled.
Details of practical applications are left for subsequent
research.

Our calculation could be simply generalized for the effects of
the gradient of the selection function. 
However, Equation~\ref{eq:legendre} could not be generalized due to
odd associated Legendre functions entering in the expansion.
These cannot be re-expressed as Legendre polynomials, thus
the final results would not admit a Legendre expansion.
This can be explicitly demonstrated from the
final results of \cite{SzalayEtal1998}. There appear to be no
analogous problems when generalizing Equation~\ref{eq:trig}: we conjecture
that trigonometric polynomial expansion will be still possible,
with terms of odd orders appearing. 
The simplicity of the present theory opens up
the possibility of generalizations for higher orders, such as three-point
correlation functions and cumulant correlators. These calculations
will be presented elsewhere.

It is a pleasure to thank Alex Szalay for stimulating discussions. 
The author was supported by NASA through AISR 
NAG5-11996, and ATP NASA NAG5-12101 as well as by
NSF grants AST02-06243 and ITR 1120201-128440.




\input ms.bbl

\end{document}

%% file: aas_journals.tex
\def\apj{{ApJ}}                 
\def\apjl{{ApJ}}                
\def\mnras{{MNRAS}}             
\def\nat{{Nature}}              



%% file: ms_pp.bbl
\begin{thebibliography}{}

\bibitem[\protect\citeauthoryear{{Bond}, {Jaffe}, \& {Knox}}{{Bond}
  et~al.}{2000}]{BondEtal2000}
{Bond}, J.~R., {Jaffe}, A.~H.,  \& {Knox}, L. 2000, \apj, 533, 19

\bibitem[\protect\citeauthoryear{{Cole}, {Fisher}, \& {Weinberg}}{{Cole}
  et~al.}{1995}]{ColeEtal1995}
{Cole}, S., {Fisher}, K.~B.,  \& {Weinberg}, D.~H. 1995, \mnras, 275, 515

\bibitem[\protect\citeauthoryear{{Davis} \& {Peebles}}{{Davis} \&
  {Peebles}}{1983}]{DavisPeebles1983}
{Davis}, M.,  \& {Peebles}, P.~J.~E. 1983, \apj, 267, 465

\bibitem[\protect\citeauthoryear{{Hamilton}}{{Hamilton}}{1993}]{Hamilton1993}
{Hamilton}, A.~J.~S. 1993, \apj, 417, 19

\bibitem[\protect\citeauthoryear{{Hamilton}}{{Hamilton}}{1998}]{Hamilton1998}
{Hamilton}, A.~J.~S. 1998, in ASSL Vol. 231: The Evolving Universe, 185

\bibitem[\protect\citeauthoryear{{Hamilton} \& {Culhane}}{{Hamilton} \&
  {Culhane}}{1996}]{HamiltonCulhane1996}
{Hamilton}, A.~J.~S.,  \& {Culhane}, M. 1996, \mnras, 278, 73

\bibitem[\protect\citeauthoryear{{Hawkins} et~al.}{{Hawkins}
  et~al.}{2003}]{HawkinsEtal2003}
{Hawkins}, E., et~al. 2003, \mnras, 346, 78

\bibitem[\protect\citeauthoryear{{Heavens} \& {Taylor}}{{Heavens} \&
  {Taylor}}{1995}]{HeavensTaylor1995}
{Heavens}, A.~F.,  \& {Taylor}, A.~N. 1995, \mnras, 275, 483

\bibitem[\protect\citeauthoryear{{Kaiser}}{{Kaiser}}{1987}]{Kaiser1987}
{Kaiser}, N. 1987, \mnras, 227, 1

\bibitem[\protect\citeauthoryear{{Peacock} et~al.}{{Peacock}
  et~al.}{2001}]{PeacockEtal2001}
{Peacock}, J.~A., et~al. 2001, \nat, 410, 169

\bibitem[\protect\citeauthoryear{{Peebles}}{{Peebles}}{1980}]{Peebles1980}
{Peebles}, P.~J.~E. 1980, {The large-scale structure of the universe} (Research
  supported by the National Science Foundation.~Princeto n, N.J., Princeton
  University Press, 1980.~435 p.)

\bibitem[\protect\citeauthoryear{{Pope} et~al.}{{Pope}
  et~al.}{2004}]{PopeEtal2004}
{Pope}, A.~C., et~al. 2004, ArXiv Astrophysics e-prints

\bibitem[\protect\citeauthoryear{{Scoccimarro}, {Couchman}, \&
  {Frieman}}{{Scoccimarro} et~al.}{1999}]{ScoccimarroEtal1999}
{Scoccimarro}, R., {Couchman}, H.~M.~P.,  \& {Frieman}, J.~A. 1999, \apj, 517,
  531

\bibitem[\protect\citeauthoryear{{Szalay}, {Matsubara}, \& {Landy}}{{Szalay}
  et~al.}{1998}]{SzalayEtal1998}
{Szalay}, A.~S., {Matsubara}, T.,  \& {Landy}, S.~D. 1998, \apjl, 498, L1

\bibitem[\protect\citeauthoryear{{Szapudi}, {Prunet}, \& {Colombi}}{{Szapudi}
  et~al.}{2001}]{SzapudiEtal2001}
{Szapudi}, I., {Prunet}, S.,  \& {Colombi}, S. 2001, \apjl, 561, L11

\bibitem[\protect\citeauthoryear{{Tadros} et~al.}{{Tadros}
  et~al.}{1999}]{TadrosEtal1999}
{Tadros}, H., et~al. 1999, \mnras, 305, 527

\bibitem[\protect\citeauthoryear{{Tegmark} et~al.}{{Tegmark}
  et~al.}{2003a}]{TegmarEtal2003a}
{Tegmark}, M., et~al. 2003a, ArXiv Astrophysics e-prints

\bibitem[\protect\citeauthoryear{{Tegmark} et~al.}{{Tegmark}
  et~al.}{1998}]{TegmarkEtal1998}
{Tegmark}, M., {Hamilton}, A.~J.~S., {Strauss}, M.~A., {Vogeley}, M.~S.,  \&
  {Szalay}, A.~S. 1998, \apj, 499, 555

\bibitem[\protect\citeauthoryear{{Tegmark}, {Hamilton}, \& {Xu}}{{Tegmark}
  et~al.}{2002}]{TegmarkEtal2002}
{Tegmark}, M., {Hamilton}, A.~J.~S.,  \& {Xu}, Y. 2002, \mnras, 335, 887

\bibitem[\protect\citeauthoryear{{Tegmark} et~al.}{{Tegmark}
  et~al.}{2003b}]{TegmarEtal2003b}
{Tegmark}, M., et~al. 2003b, ArXiv Astrophysics e-prints

\bibitem[\protect\citeauthoryear{{Varshalovich}, {Moskalev}, \&
  {Khershonski}}{{Varshalovich} et~al.}{1988}]{VarshalovichEtal1988}
{Varshalovich}, D.~A., {Moskalev}, A.~N.,  \& {Khershonski}, V.~K. 1988,
  {Quantum Theory of the Angular Momentum}

\bibitem[\protect\citeauthoryear{{Vogeley} \& {Szalay}}{{Vogeley} \&
  {Szalay}}{1996}]{VogelleySzalay1996}
{Vogeley}, M.~S.,  \& {Szalay}, A.~S. 1996, \apj, 465, 34

\bibitem[\protect\citeauthoryear{{Zaroubi} \& {Hoffman}}{{Zaroubi} \&
  {Hoffman}}{1996}]{ZaroubiHoffman1996}
{Zaroubi}, S.,  \& {Hoffman}, Y. 1996, \apj, 462, 25

\bibitem[\protect\citeauthoryear{{Zehavi} et~al.}{{Zehavi}
  et~al.}{2002}]{ZehaviEtal2002}
{Zehavi}, I., et~al. 2002, \apj, 571, 172

\end{thebibliography}
